# The locality of quantum subsystems III: A 'subsystem compatible' interpretation of quantum mechanics

Adam Brownstein[1]

[1]*University of Melbourne, School of Physics*

23/08/2017

A local description of quantum subsystems can be used to construct ontologies of the full quantum predictions. This paper communicates one possible way to do so. A retrocausal interpretation of quantum mechanics where the de Broglie-Bohm particles are the retrocausal agents is developed. This interpretation is constrained to be compatible with the existence of a local realist description of quantum subsystems.

# Contents

# 1 Introduction

The problem of correlations in quantum mechanical joint probability distributions has caused a profound change in our understanding of reality. Bell's theorem [1] has proven that these correlations are incompatible with simple forms of local-realism. The veracity of this theorem has been demonstrated in several experimental loophole-free Bell tests [2, 3, 4]. A very narrow selection of philosophical loopholes remain open.

Bell's theorem depends upon several (quite reasonable) assumptions. First, the theorem assumes that measurements made in distant laboratories actually occur. This is a form of counterfactual definiteness about the quantum mechanical joint probabilities which could be denied in principle. Second, the theorem assumes that local, subluminal causal influences travel within forward light cones. Therefore any non-classical correlations between distant laboratories are believed to have propagated either superluminally or non-locally. However, retrocausal effects - which propagate within backward light cones - can evade this conclusion. Third, the theorem assumes that the choices of measurement basis of each entangled particle are not predetermined in a conspiratorial way. Rejecting this assumption is the approach of superdeterminism.

This paper will incorporate the local interpretation of quantum marginals developed in papers one [5] and two [6] of this series into an interpretation of the full quantum predictions. But to explain the quantum mechanical joint probabilities in a local way, it is necessary to evade the penetrating conclusions of Bell's theorem. We exploit the retrocausality loophole in this theorem and establish a modified de Broglie-Bohm interpretation of quantum mechanics where particles are the retrocausal agents. The resulting interpretation demonstrates that describing the quantum mechanical joint probabilities does not have to come at the expense of the description of the marginal probabilities. The interpretation contains a local, forwardly causal description of the marginal probabilities embedded within a broader framework for the joint probabilities. We have described the interpretation as 'subsystem compatible' since it is compatible with the existence of a local realist description of the marginal probabilities.

To establish this *Subsystem Compatible* interpretation, Section 2 first develops a local de Broglie-Bohm interpretation of single-particle subsystems. This is enabled by a local sum-over-paths decomposition of the probability current density of the subsystem. As well as being central to our broader aims, the results here stand as an independent contribution to the single-particle case of the de Broglie-Bohm interpretation. Section 3 then develops the main thesis of this Subsystem Compatible approach.

# 2 A local de Broglie-Bohm theory for quantum subsystems

The aim of this section is to construct a local de Broglie-Bohm interpretation of single-particle subsystems. This is intended to achieve two objectives. A local interpretation of single-particle subsystems will later enable an interpretation of the total quantum system. But also, the result provides a standalone contribution to the de Broglie-Bohm interpretation. The standard single-particle de Broglie-Bohm interpretation is important in isolation, since it provides an intuitive and largely paradox free[1] description of quantum mechanics its restricted domain of validity. We wish to extend this domain of validity to account for single-particle subsystems which are entangled to external systems.

To construct local guidance equations for particles of the subsystem, the probability density and probability current density of the subsystem are required. To obtain these two quantities, we apply a local sum-over-paths decomposition to Dirac current of the subsystem. The Dirac current is useful for this task since it is a bilinear observable constructed directly from the wavefunction of the Dirac particle. Consequently, it describes the probability current density without the use of derivatives, and hence is amenable to local sum-over-paths decomposition.

---

[1] See for example, Einstein's comparison of de Broglie's interpretation and the Copenhagen interpretation for a single-slit experiment in Bacciagaluppi & Valentini [7].

Since the Dirac current intrinsically depends on spin degrees of freedom, our previous methods [6] need to be generalized to the case of both spatial and spin entanglement, which is done in Appendix A. Section 2.1 obtains the Dirac current of the subsystem, while sections 2.2 and 2.3 recast this current in terms of the local sum-over-paths ontology. Readers primarily interested in the Subsystem Compatible interpretation may wish to read Section 2.3 only, before moving to Section 3.

## 2.1 Two-body Dirac current

Let $\Psi(\vec{x}_1,\vec{x}_2,t)$ be the wavefunction of the two-body Dirac system. This wavefunction is a 16 component vector, with components $\Psi_{\alpha\beta}(\vec{x}_1,\vec{x}_2,t)$ where $\alpha$ and $\beta$ are spinor indices.[2] The Schrodinger equation for the two-body Dirac wavefunction is:

$$i\hbar \frac{\partial \Psi(\vec{x}_1,\vec{x}_2,t)}{\partial t} = H\Psi(\vec{x}_1,\vec{x}_2,t), \tag{2.1}$$

where the Hamiltonian $H = H_1 + H_2$ such that:

$$H_1 = \left[-i\hbar \gamma^0 \gamma^j \left(\partial_j - iqA_j(\vec{x}_1,t)\right) + (m + qA_0(\vec{x}_1,t))\gamma^0\right] \otimes \mathbb{I}, \tag{2.2}$$

$$H_2 = \mathbb{I} \otimes \left[-i\hbar \gamma^0 \gamma^j \left(\partial_j - iqA_j(\vec{x}_2,t)\right) + (m + qA_0(\vec{x}_2,t))\gamma^0\right]. \tag{2.3}$$

The gauge potentials $A_\mu(\vec{x}_1,t)$ and $A_\mu(\vec{x}_2,t)$ are assumed to be classical electromagnetic four potentials, since we are working in a semiclassical framework. The quantity $|\Psi(\vec{x}_1,\vec{x}_2,t)|^2 = \Psi^\dagger(\vec{x}_1,\vec{x}_2)\Psi(\vec{x}_1,\vec{x}_2)$ is normalized and positive definite, and hence can be regarded as the probability for measuring particles one and two at positions $\vec{x}_1$ and $\vec{x}_2$ respectively. A continuity equation for this probability density can be derived in the following way:

$$\begin{aligned}\frac{\partial |\Psi(\vec{x}_1,\vec{x}_2,t)|^2}{\partial t} &= \frac{\partial \Psi^\dagger(\vec{x}_1,\vec{x}_2)\Psi(\vec{x}_1,\vec{x}_2)}{\partial t} \\ &= \frac{\partial \Psi^\dagger(\vec{x}_1,\vec{x}_2,t)}{\partial t}\Psi(\vec{x}_1,\vec{x}_2,t) + \Psi^\dagger(\vec{x}_1,\vec{x}_2,t)\frac{\partial \Psi(\vec{x}_1,\vec{x}_2,t)}{\partial t} \\ &= \frac{i}{\hbar}\left(-i\hbar \frac{\partial \Psi^\dagger(\vec{x}_1,\vec{x}_2,t)}{\partial t}\right)\Psi(\vec{x}_1,\vec{x}_2) + \frac{-i}{\hbar}\Psi^\dagger(\vec{x}_1,\vec{x}_2,t)\left(i\hbar\frac{\partial \Psi(\vec{x}_1,\vec{x}_2,t)}{\partial t}\right) \quad (2.4) \\ &= \frac{i}{\hbar}(H\Psi(\vec{x}_1,\vec{x}_2,t))^\dagger \Psi(\vec{x}_1,\vec{x}_2,t) + \frac{-i}{\hbar}\Psi^\dagger(\vec{x}_1,\vec{x}_2,t)(H\Psi(\vec{x}_1,\vec{x}_2,t)) \\ &= \frac{i}{\hbar}(H_1\Psi(\vec{x}_1,\vec{x}_2,t))^\dagger \Psi(\vec{x}_1,\vec{x}_2,t) + \frac{-i}{\hbar}\Psi^\dagger(\vec{x}_1,\vec{x}_2,t)(H_1\Psi(\vec{x}_1,\vec{x}_2,t)) \\ &\quad + \frac{i}{\hbar}(H_2\Psi(\vec{x}_1,\vec{x}_2,t))^\dagger \Psi(\vec{x}_1,\vec{x}_2,t) + \frac{-i}{\hbar}\Psi^\dagger(\vec{x}_1,\vec{x}_2,t)(H_2\Psi(\vec{x}_1,\vec{x}_2,t)) \quad (2.5) \\ &= -\partial_i j^i_{\vec{x}_1}(\vec{x}_1,\vec{x}_2,t) - \partial_i j^i_{\vec{x}_2}(\vec{x}_1,\vec{x}_2), \quad (2.6)\end{aligned}$$

where we used the Schrodinger equation and its conjugate in line (2.4) and have substituted $H = H_1 + H_2$. The linearity of the adjoint then gives Eq. (2.5). The explicit expressions for $H_1$ and $H_2$ are subsequently inserted to derive Eq. (2.6). Note that because $H_1$ and $H_2$ are essentially single particle Dirac Hamiltonians, the derivation of Eq. (2.6) from Eq. (2.5) carries through analogous to the derivation of a single-particle Dirac current from a Schrodinger equation. The currents $j^i_{\vec{x}_1}(\vec{x}_1,\vec{x}_2,t)$ and $j^i_{\vec{x}_2}(\vec{x}_1,\vec{x}_2)$ in this expression are defined as:

$$j^i_{\vec{x}_1}(\vec{x}_1,\vec{x}_2,t) \equiv \Psi(\vec{x}_1,\vec{x}_2,t)^\dagger \left(\gamma^0 \gamma^i \otimes \mathbb{I}\right) \Psi(\vec{x}_1,\vec{x}_2,t). \qquad j^i_{\vec{x}_2}(\vec{x}_1,\vec{x}_2) \equiv \Psi(\vec{x}_1,\vec{x}_2,t)^\dagger \left(\mathbb{I} \otimes \gamma^0 \gamma^i\right) \Psi(\vec{x}_1,\vec{x}_2,t). \tag{2.7}$$

---

[2] These spinor indices allow spin entanglement to be encoded between Dirac particles.

Bringing the currents to the left-hand side in Eq. (2.6) then produces the continuity equation:

$$\frac{\partial |\Psi(\vec{x}_1,\vec{x}_2,t)|^2}{\partial t} + \frac{\partial}{\partial x_1^i} j_{\vec{x}_1}^i(\vec{x}_1,\vec{x}_2,t) + \frac{\partial}{\partial x_2^i} j_{\vec{x}_2}^i(\vec{x}_1,\vec{x}_2,t) = 0. \tag{2.8}$$

By integrating with respect to the coordinates of particle two, the divergence theorem can be used to eliminate the contributions from $j_{\vec{x}_2}^i(\vec{x}_1,\vec{x}_2,t)$, giving the continuity equation for particle ones's subsystem:

$$\begin{aligned} 0 &= \frac{\partial}{\partial t}\left(\int j^0(\vec{x}_1,\vec{x}_2,t)\,d\vec{x}_2\right) + \frac{\partial}{\partial x_1^i}\int j_{\vec{x}_1}^i(\vec{x}_1,\vec{x}_2,t)\,d\vec{x}_2 + \int \frac{\partial}{\partial x_2^i} j_{\vec{x}_2}^i(\vec{x}_1,\vec{x}_2,t)\,d\vec{x}_2, \\ &= \frac{\partial}{\partial t}\left(j^0(\vec{x}_1,t)\right) + \frac{\partial}{\partial x_1^i}\mathbf{j}(\vec{x}_1,t), \end{aligned} \tag{2.9}$$

where we have defined:

$$j^0(\vec{x}_1,t) \equiv \int j_{\vec{x}_1}^0(\vec{x}_1,\vec{x}_2,t)\,d\vec{x}_2. \qquad \mathbf{j}(\vec{x}_1,t) \equiv \int \mathbf{j}_{\vec{x}_1}(\vec{x}_1,\vec{x}_2,t)\,d\vec{x}_2. \tag{2.10}$$

## 2.2 Local sum-over-paths for the Dirac current

A local hidden variable interpretation for $j^\mu(\vec{x}_1,t) \equiv \int j_{\vec{x}_1}^\mu(\vec{x}_1,\vec{x}_2,t)\,d\vec{x}_2$ can now be obtained in the following manner. First, write out the Dirac current for the subsystem, then convert the matrix notation into an explicit summation over spinor indices:

$$\begin{aligned} j^\mu(\vec{x}_1,t) &= \int d\vec{x}_2 \Psi(\vec{x}_1,\vec{x}_2,t)^\dagger \left(\gamma^0\gamma^\mu \otimes \mathbb{I}\right) \Psi(\vec{x}_1,\vec{x}_2,t) \\ &= \int d\vec{x}_2 \sum_a\sum_b\sum_c\sum_d \left(\Psi_{(a,b)}(\vec{x}_1,\vec{x}_2,t)\right)^\dagger \gamma^0_{(a,d)}\gamma^\mu_{(d,c)}\Psi_{(c,b)}(\vec{x}_1,\vec{x}_2,t) \\ &= \int d\vec{x}_2 \sum_a\sum_b\sum_c \left(\Psi_{(a,b)}(\vec{x}_1,\vec{x}_2,t)\right)^\dagger \gamma^0_{(a,a)}\gamma^\mu_{(a,c)}\Psi_{(c,b)}(\vec{x}_1,\vec{x}_2,t). \end{aligned} \tag{2.11}$$

We have removed the sum over $d$ since $\gamma^0$ is a diagonal matrix in the Dirac representation of the gamma matrices, so $\gamma^0_{(a,d)}$ is non-zero only for $a = d$. The components of the wavefunction can be expressed as $\Psi_{(a,b)}(\vec{x}_1,\vec{x}_2,t) = \langle \vec{x}_1,a,\vec{x}_2,b|\Psi(t)\rangle$, where $|\Psi(t)\rangle$ is a ray in the Hilbert space for the two-body Dirac system. Then using $|\Psi(t)\rangle = \mathbb{U}(t)|\Psi(0)\rangle$ gives $\Psi_{(a,b)}(\vec{x}_1,\vec{x}_2,t) = \langle \vec{x}_1,a,\vec{x}_2,b|\mathbb{U}(t)|0\rangle$, where $\mathbb{U}(t)$ is a unitary operator acting on kets from the Hilbert space $\mathcal{H}_{\text{Space A}} \otimes \mathcal{H}_{\text{Spin A}} \otimes \mathcal{H}_{\text{Space B}} \otimes \mathcal{H}_{\text{Spin B}}$. Consequently, the current becomes:

$$\begin{aligned} j^\mu(\vec{x}_1,t) &= \int d\vec{x}_2 \sum_a\sum_b\sum_c \langle \Psi(0)|\mathbb{U}(t)^\dagger|\vec{x}_1,a,\vec{x}_2,b\rangle \gamma^0_{(a,a)}\gamma^\mu_{(a,c)} \langle \vec{x}_1,c,\vec{x}_2,b|\mathbb{U}(t)|\Psi(0)\rangle \\ &= \int d\vec{x}_2 \sum_a\sum_b\sum_c \gamma^0_{(a,a)}\gamma^\mu_{(a,c)} \langle \Psi(0)|\mathbb{U}(t)^\dagger|\vec{x}_1,a,\vec{x}_2,b\rangle \langle \vec{x}_1,c,\vec{x}_2,b|\mathbb{U}(t)|\Psi(0)\rangle. \end{aligned} \tag{2.12}$$

Appendix A demonstrates how to generalize the local sum-over-paths interpretation to space×spin paths. Condition the unitary matrices on the space×spin paths of particle one. Explicitly, we have:

$$\langle \Psi(0)|\mathbb{U}(t)^\dagger|\vec{x}_1,a,\vec{x}_2,b\rangle = \sum_{P^{(\vec{x}_1,a,t)}} A_P \langle \Psi(0)|\mathbb{U}_P(t)^\dagger|\vec{x}_2,b\rangle, \tag{2.13}$$

$$\langle \vec{x}_1,c,\vec{x}_2,b|\mathbb{U}(t)|\Psi(0)\rangle = \sum_{Q^{(\vec{x}_1,c,t)}} A_Q \langle \vec{x}_2,b|\mathbb{U}_Q(t)|\Psi(0)\rangle, \tag{2.14}$$

where $\mathbb{U}_P(t)$ and $\mathbb{U}_Q(t)$ are the unitary operator for particle two given that particle one traverses paths $P$ and $Q$ respectively. Therefore:

$$j^\mu(\vec{x}_1,t) = \int d\vec{x}_2 \sum_a \sum_b \sum_c \gamma^0_{(a,a)} \gamma^\mu_{(a,c)} \langle\Psi(0)|\mathbb{U}(t)^\dagger|\vec{x}_1,a,\vec{x}_2,b\rangle\langle\vec{x}_1,c,\vec{x}_2,b|\mathbb{U}(t)|\Psi(0)\rangle$$

$$= \int d\vec{x}_2 \sum_a \sum_b \sum_c \sum_{P(\vec{x}_1,a,t)} \sum_{Q(\vec{x}_1,c,t)} \gamma^0_{(a,a)} \gamma^\mu_{(a,c)} A_P^* A_Q \langle\Psi(0)|\mathbb{U}_P(t)^\dagger|\vec{x}_2,b\rangle\langle\vec{x}_2,b|\mathbb{U}_Q(t)|\Psi(0)\rangle$$

$$= \sum_a \sum_c \sum_{P(\vec{x}_1,a,t)} \sum_{Q(\vec{x}_1,c,t)} \gamma^0_{(a,a)} \gamma^\mu_{(a,c)} A_P^* A_Q \langle\Psi(0)|\mathbb{U}_P(t)^\dagger \left(\int d\vec{x}_2 \sum_b |\vec{x}_2,b\rangle\langle\vec{x}_2,b|\right) \mathbb{U}_Q(t)|\Psi(0)\rangle$$

$$= \sum_a \sum_c \sum_{P(\vec{x}_1,a,t)} \sum_{Q(\vec{x}_1,c,t)} \gamma^0_{(a,a)} \gamma^\mu_{(a,c)} A_P^* A_Q \langle\Psi(0)|\mathbb{U}_P(t)^\dagger \mathbb{U}_Q(t)|\Psi(0)\rangle, \qquad (2.15)$$

where we have used the completeness relation $\int d\vec{x}_2 \sum_b |\vec{x}_2,b\rangle\langle\vec{x}_2,b| = \int d\vec{x}_2 |\vec{x}_2\rangle\langle\vec{x}_2| \otimes \sum_b |b\rangle\langle b| = \mathbb{I} \otimes \mathbb{I}$. Having performed these manipulations, we are now ready to apply a local sum-over-paths decomposition to the term $\langle\Psi(0)|\mathbb{U}_P(t)^\dagger \mathbb{U}_Q(t)|\Psi(0)\rangle$, which appears in Eq. (2.15). The result of this decomposition will be:

$$\langle\Psi(0)|\mathbb{U}_P(t)^\dagger \mathbb{U}_Q(t)|\Psi(0)\rangle = 1 + \sum_{r=1}^t H_{P,Q}^{(r)} \equiv \lambda_{P,Q}^{(t)}, \qquad (2.16)$$

where $H_{P,Q}^{(r)}$ are hits of information obtained from interactions, and $\lambda_{P,Q}$ are hidden variables which control the interference occurring between path amplitudes $A_P$ and $A_Q$. Explicitly, we have:

$$j^\mu(\vec{x}_1,t) = \sum_a \sum_c \sum_{P(\vec{x}_1,a,t)} \sum_{Q(\vec{x}_1,c,t)} \gamma^0_{(a,a)} \gamma^\mu_{(a,c)} A_P^* A_Q \lambda_{P,Q}^{(t)}, \qquad (2.17)$$

To explore this expression further, set $\mu = 0$. Eq. (2.17) reduces to:

$$j^0(\vec{x}_1,t) = \sum_a \sum_c \sum_{P(\vec{x}_1,a,t)} \sum_{Q(\vec{x}_1,c,t)} \gamma^0_{(a,a)} \gamma^0_{(a,c)} A_P^* A_Q \lambda_{P,Q}^{(t)}$$

$$= \sum_a \sum_c \sum_{P(\vec{x}_1,a,t)} \sum_{Q(\vec{x}_1,c,t)} \gamma^0_{(a,a)} \gamma^0_{(a,a)} A_P^* A_Q \lambda_{P,Q}^{(t)} \qquad (2.18)$$

$$= \sum_a \sum_{P(\vec{x}_1,a,t)} \sum_{Q(\vec{x}_1,c,t)} A_P^* A_Q \lambda_{P,Q}^{(t)}$$

$$= \sum_a \left( \sum_{P(\vec{x}_1,a,t)} |A_P|^2 + \sum_{P(\vec{x}_1,a,t) \neq Q(\vec{x}_1,a,t)} A_P^* A_Q \lambda_{P,Q}^{(t)} \right), \qquad (2.19)$$

where we have eliminated the sum over c in line (2.18) since $\gamma^0_{(a,c)} = 0$ if $a \neq c$. The relation $\gamma^0_{(a,a)} \gamma^0_{(a,a)} = 1$ has then been used to eliminate the gamma matrix elements from this expression. The resulting Eq. (2.19) should be familiar as a local sum-over-paths for the probability density of the subsystem. The only real difference to the two-particle case of paper two [6] is that the sum is over space×spin paths. Now for $\mu = 1,2,3$ we have:

$$j^i(\vec{x}_1,t) = \sum_a \sum_c \sum_{P(\vec{x}_1,a,t)} \sum_{Q(\vec{x}_1,c,t)} \gamma^0_{(a,a)} \gamma^j_{(a,c)} A_P^* A_Q \lambda_{P,Q}^{(t)}$$

$$= \left( \sum_a \gamma^0_{(a,a)} \gamma^j_{(a,a)} \sum_{P(\vec{x}_1,a,t)} |A_P|^2 + \sum_a \sum_c \sum_{P(\vec{x}_1,a,t) \neq Q(\vec{x}_1,a,t)} \gamma^0_{(a,a)} \gamma^j_{(a,c)} A_P^* A_Q \lambda_{P,Q}^{(t)} \right) \qquad (2.20)$$

$$= \sum_a \sum_c \sum_{P(\vec{x}_1,a,t) \neq Q(\vec{x}_1,c,t)} \gamma^0_{(a,a)} \gamma^j_{(a,c)} A_P^* A_Q \lambda_{P,Q}^{(t)}. \qquad (2.21)$$

The first term in Eq. (2.20) disappears, because $\gamma^i_{(a,a)} = 0$ for $i = \in \{1,2,3\}$, assuming the Dirac representation of the gamma matrices.

## 2.3 Local guidance equations for the subsystem

Having provided a local interpretation of the probability density Eq. (2.19) and probability current density Eq. (2.21) of the subsystem, a local de Broglie-Bohm interpretation of the subsystem is immediately available. Suppose that particles of the subsystem are initialized in a statistical ensemble which matches the distribution $j^0(\vec{x}_1, 0)$. Then if their trajectories are determined by the following guidance equation:

$$\frac{d\mathbf{X}_1(t)}{dt} = \frac{\mathbf{j}(\vec{x}_1,t)}{j^0(\vec{x}_1,t)}\bigg|_{\vec{x}_1 = \mathbf{X}_1(t)}, \quad (2.22)$$

by the uniqueness of solutions to continuity equations, they will remain distributed as $j^0(\vec{x}_1, t)$ for all subsequent times.[3] Recall that $j^0(\vec{x}_1,t) = \int |\Psi(\vec{x}_1,\vec{x}_2,t)|^2 d\vec{x}_2$ is the marginal probability density for particle one, therefore a statistical ensemble of de Broglie-Bohm particles will match the quantum predictions for this particle. This de Broglie-Bohm interpretation will be local if the particles sample the components of the currents $j^0(\vec{x}_1,t)$ and $j^i(\vec{x}_1,t)$ in accordance with the local sum-over-paths interpretation.

## 2.4 Discretized space or spatial modes?

Although a local sum-over-paths decomposition can be performed by dividing the continuous spatial coordinates into a lattice of discrete positions, the ontology is quite complex due to the large number of paths available through the discretized position space. However, there is an alternative method worth considering.

Suppose that fundamentally, only wavepackets can become entangled, not the individual positions within wavepackets. In Hilbert space, these wavepackets correspond to spatial modes. We can construct a sum-over-paths interpretation based on entanglement between spatial modes by inserting the completeness relations $\sum_j |j\rangle_A \langle j|_A = \mathbb{I}$ and $\sum_q |q\rangle_A \langle q|_A = \mathbb{I}$ into the expression for the probability current density. Doing this explicitly:

$$\begin{aligned}
&j^\mu(\vec{x}_1, t) \\
&= \int d\vec{x}_2 \sum_a \sum_b \sum_c \gamma^0_{(a,a)} \gamma^\mu_{(a,c)} \langle \Psi(0)|\mathbb{U}(t)^\dagger |\vec{x}_1, a, \vec{x}_2, b\rangle \langle \vec{x}_1, c, \vec{x}_2, b|\mathbb{U}(t)|\Psi(0)\rangle \\
&= \int d\vec{x}_2 \sum_a \sum_b \sum_c \gamma^0_{(a,a)} \gamma^\mu_{(a,c)} \langle \Psi(0)|\mathbb{U}(t)^\dagger \left(\sum_j |j\rangle_A \langle j|_A\right) |\vec{x}_1, a, \vec{x}_2, b\rangle \langle \vec{x}_1, c, \vec{x}_2, b| \left(\sum_q |q\rangle_A \langle q|_A\right) \mathbb{U}(t)|\Psi(0)\rangle \\
&= \int d\vec{x}_2 \sum_a \sum_b \sum_c \sum_j \sum_q \Big(\gamma^0_{(a,a)} \gamma^\mu_{(a,c)} \langle j|\vec{x}_1\rangle \langle \vec{x}_1|q\rangle ... \\
&\quad ... \times \langle \Psi(0)|\mathbb{U}(t)^\dagger |j, a, \vec{x}_2, b\rangle \langle q, c, \vec{x}_2, b|\mathbb{U}(t)|\Psi(0)\rangle \Big)
\end{aligned} \quad (2.23)$$

The terms $\langle \vec{x}_1|j\rangle$ and $\langle \vec{x}_1|q\rangle$ are probability amplitudes for individual wavepackets of particle A.

---

[3] This property is known as equivariance.

Rather than condition upon space×spin paths, we condition upon spatial mode×spin paths:

$$j^\mu(\vec{x}_1, t) = \int d\vec{x}_2 \sum_a \sum_b \sum_c \sum_j \sum_q \left( \gamma^0_{(a,a)} \gamma^\mu_{ac} \langle j|\vec{x}_1\rangle \langle \vec{x}_1|q\rangle \ldots \right.$$

$$\left. \ldots \times \sum_{P(j,a,t)} \sum_{Q(q,c,t)} A_P^* A_Q \langle \Psi(0)|\mathbb{U}_P(t)^\dagger|\vec{x}_2, b\rangle \langle \vec{x}_2, b|\mathbb{U}_Q(t)|\Psi(0)\rangle \right)$$

$$= \sum_a \sum_c \sum_j \sum_q \left( \gamma^0_{(a,a)} \gamma^\mu_{(a,c)} \langle j|\vec{x}_1\rangle \langle \vec{x}_1|q\rangle \ldots \right.$$

$$\left. \ldots \times \sum_{P(j,a,t)} \sum_{Q(q,c,t)} A_P^* A_Q \langle \Psi(0)|\mathbb{U}_P(t)^\dagger \left( \int d\vec{x}_2 \sum_b |\vec{x}_2,b\rangle\langle \vec{x}_2,b| \right) \mathbb{U}_Q(t)|\Psi(0)\rangle \right) \quad (2.24)$$

$$= \sum_a \sum_c \sum_j \sum_q \gamma^0_{(a,a)} \gamma^\mu_{ac} \langle j|\vec{x}_1\rangle \langle \vec{x}_1|q\rangle \sum_{P(j,a,t)} \sum_{Q(q,c,t)} A_P^* A_Q \langle \Psi(0)|\mathbb{U}_P(t)^\dagger \mathbb{U}_Q(t)|\Psi(0)\rangle. \quad (2.25)$$

The completeness relation $\int d\vec{x}_2 \sum_b |\vec{x}_2, b\rangle\langle \vec{x}_2, b| = \int d\vec{x}_2 |\vec{x}_2\rangle\langle \vec{x}_2| \otimes \sum_b |b\rangle\langle b| = \mathbb{I} \otimes \mathbb{I}$ has been used in line (2.24) to produce Eq. (2.25). A local sum-over-paths interpretation is now readily available by decomposing $\langle \Psi(0)|\mathbb{U}_P(t)^\dagger \mathbb{U}_Q(t)|\Psi(0)\rangle$. The question of whether quantum mechanics fundamentally describe entanglement between wavepackets (or spatial modes), or entanglement between individual positions within wavepackets will require further theoretical and experimental determination.

## 2.5 Summary of results

This section has established a local de Broglie-Bohm interpretation of single-particle subsystems. These results extend the standard single-particle de Broglie-Bohm interpretation, which is valid only for unentangled wavefunctions, to the case where the single-particle subsystem may be entangled to an external system.

We found it convenient to use the Dirac current in the construction of guidance equations for particles of the subsystem. The local sum-over-paths decomposition can be applied to the Dirac current after conditioning upon either space×spin or spatial mode×spin paths of the subsystem. Attempting to reconcile the complex ontology of a local sums-over-paths on a discretized position space leads us to consider that spatial mode entanglement might be the fundamental form of position entanglement.

# 3 The Subsystem Compatible interpretation

In this section, we provide a first example of how a local interpretation of quantum subsystems can be incorporated into an explanation of the entire quantum predictions. We construct a retrocausal interpretation of quantum mechanics where particles are the retrocausal agents. The interpretation is made possible through the local de Broglie-Bohm interpretation of single-particle subsystems, coupled with a change in perspective on the specification of initial and final boundary conditions in the de Broglie-Bohm interpretation.

## 3.1 Retrocausal de Broglie-Bohm particles

Because the de Broglie-Bohm interpretation is a deterministic theory of particle motion, it is mostly irrelevant in which direction of time the particles are moving. A similar situation occurs in classical mechanics, where it is widely recognized that the boundary conditions specify the arrow of time rather than the dynamical laws of nature. We always observe entropy increasing in the forward direction of time, however the underlying equations of motion are time symmetric. This increase in entropy is easily misattributed as forward causation. However, in the classical context at least, the emergence of forward causation arises from ignorance about the precise microscopic conditions.

In the de Broglie-Bohm interpretation, the atemporal nature of the particle dynamics is especially evident. Because the guidance equation is a first order differential equation, the motion of the particle configuration depends only on the current arrangement of the wavefunction, and not any past or future information. Relatedly, the lack of back-reaction of the particle on the wavefunction indicates the motion of particle and wavefunction are completely causally disconnected. These features make it very easy to recast the de Broglie-Bohm interpretation as a retrocausal interpretation. This can be achieved by asserting that the de Broglie-Bohm particle configuration propagates in the opposite temporal direction to the wavefunction; and by specifying the final, rather than initial, boundary conditions for the particle configuration.

A change of boundary conditions in the de Broglie-Bohm interpretation is not widely discussed in the literature. Usually retrocausal interpretations of quantum mechanics are constructed from wavefunctions or waves propagating in opposite temporal directions, such as in Cramer's *Transactional Interpretation* [8], Aharonov and Vaidman's *Two State Vector Formalism* [9], Wharton's *Time-Symmetric Quantum Mechanics* [10] and Sutherland's *Causally Symmetric Bohm Model* [11]. While counterpropagating waves may be the correct approach, we wish to deconstruct the nature of the naive approach already embedded within the de Broglie-Bohm interpretation.

It is interesting to ask why this naive retrocausal interpretation has not been seriously examined to date. There is a very good reason. Asserting that the particle distribution is moving in the opposite temporal direction to the wavefunction does nothing to solve the real issue in the many-body de Broglie-Bohm interpretation. Even if the particles are moving backward in time, non-local guidance equations are required to make the particle distribution match the quantum mechanical joint probability distribution. So although the resulting interpretation will be retrocausal, it will remain non-local.

Nevertheless, we highlight that there is a way to recover a local retrocausal interpretation based on retrocausal de Broglie-Bohm particles. Rather than use the many-body guidance equations to propagate the particle distribution backward in time, use the de Broglie-Bohm guidance equations of single-particle subsystems. This raises the question of what happens to the correlations which were imprinted into the particle configuration at the choice of the final boundary condition. Although the particle configuration is propagated according to the wrong guidance equations, the majority of this correlation will be preserved, and the preserved elements of the correlation are sufficient to explain the quantum predictions.

### 3.2 The Subsystem Compatible interpretation (retrocausal version)

We now commence a formal presentation of the *Subsystem Compatible* interpretation. In the Subsystem Compatible interpretation, we make four simple postulates.

1. First, we assert that the configuration space wavefunction[4] evolves in the forward direction of time according to a Schrodinger equation:

$$i\hbar \frac{\partial \Psi(\vec{x}_1, \vec{x}_1, ..., \vec{x}_n, t)}{\partial t} = H \Psi(\vec{x}_1, \vec{x}_1, ..., \vec{x}_n, t). \tag{3.1}$$

2. Second, we assert that there is an actual de Broglie-Bohm particle configuration $\vec{X}(t) = (X_1(t), ..., X_n(t))$ which is localized at all times during the time evolution. This actual particle configuration $\vec{X}(t)$ is comprised of a collection of actual particle positions $X_i(t)$.

3. Third, we assert that the particle configuration is specified at the final time of the universe $t_f$, as a final boundary condition. The particle configuration is chosen to be a random sample from the joint probability distribution $|\Psi(\vec{x}_1, \vec{x}_1, ..., \vec{x}_n, t_f)|^2$ provided by the configuration space wavefunction.

---

[4]The evolution of the configuration space wavefunction can also be embedded into three-dimensional space by various means, for instance using a traditional sum-over-paths interpretation.

4. Fourth, we assert that individual particles $\mathbf{X}_i(t)$ evolve in the reverse direction of time according to the de Broglie-Bohm guidance equations of single-particle subsystems. Explicitly, we have:

$$\frac{d\mathbf{X}_i(t)}{dt} = \left.\frac{\mathbf{j}(\vec{x}_i,t)}{|\Psi(x_i,t)|^2}\right|_{\vec{x}_i = \mathbf{X}_i(t)}, \quad (3.2)$$

for each particle $i$. In this expression, $|\Psi(x_1,t)|$ is the marginal probability density of the subsystem:

$$|\Psi(x_i,t)|^2 = \int |\Psi(\vec{x}_1,...,\vec{x}_n,t)|^2 d\vec{x}_1...d\vec{x}_{i-1}d\vec{x}_{i+1}d\vec{x}_n. \quad (3.3)$$

while the term $\mathbf{j}(\vec{x}_i,t)$ is the probability current density of the subsystem:

$$\mathbf{j}(\vec{x}_i,t) = \int \mathbf{j}_{\vec{x}_i}(\vec{x}_1,...,\vec{x}_n,t) d\vec{x}_1...d\vec{x}_{i-1}d\vec{x}_{i+1}d\vec{x}_n. \quad (3.4)$$

The guidance equations (Eq. (3.2)) are local if they are described within a local ontology, for instance the local sum-over-paths approach of Section 2.

The above postulates are sufficient to reproduce the quantum predictions for all observable phenomena.

## 3.3 Reproducing the quantum predictions

Since the particles are chosen in accordance with the final joint probability distribution $|\Psi(\vec{x}_1, \vec{x}_1, ..., \vec{x}_n, t_f)|^2$ (postulate 3), we know that each individual particle is in agreement with its particular marginal probability distribution $|\Psi(\vec{x}_i, t_f)|^2$. Consequently, as the particles are guided according to the guidance equations of single-particle subsystems (postulate 4), they must remain in agreement with their respective marginal distributions $|\Psi(\vec{x}_i, t)|^2$ for all earlier times. This is due to the equivariance property of the de Broglie-Bohm interpretation of single-particle subsystems.

The essence of the Subsystem Compatible interpretation is that an ensemble of final particle configurations, propagating backward in time according to the de Broglie-Bohm guidance equations for single-particle subsystems, form a branching topological structure in configuration space. This topological structure, for the most part, matches the topological structure formed by an ensemble of particle configurations propagating in the standard de Broglie-Bohm interpretation. The common elements between the two branching structures correspond to the entire range of observable phenomena of quantum mechanics. We will now contrast the guidance of particle configurations in the standard de Broglie-Bohm interpretation and the Subsystem Compatible interpretation to provide more detail on this approach.

**Standard de Broglie-Bohm interpretation.** The time evolution in the standard de Broglie-Bohm interpretation has the following properties:

1. In the single-particle de Broglie-Bohm interpretation, the measurement process consists of wavepackets being induced into regions of disjoint spatial support in one-to-one correspondence with the eigenstates of the measurement operator being applied to the system [12, 13]. The proportion of particles of a statistical ensemble in each of the output wavepackets corresponds to the measurement probabilities.

2. Analogously, in the many-body de Broglie-Bohm interpretation, the measurement process consists of configuration space wavepackets being induced into regions of disjoint spatial support in configuration space. The proportion of particle configurations of a statistical ensemble in each of the output configuration space wavepackets corresponds to the measurement probabilities.

3. If an ensemble of particle configurations is watched over time, the trajectories of these particle configurations form a branching topological structure in configuration spacetime (i.e. a branching structure in the combined space of configurations and the temporal dimension). This branching structure describes a collection of worlds in an analogous sense to the Many-Worlds interpretation; but at the probability level rather than the wavefunction amplitude level. We will call this branching structure *Track-1* for future reference.

4. The particle configuration ends up associated with one of the configuration space wavepackets (or alternatively, end up in a particular branch of Track-1). In other words, the configuration becomes localized within a single world upon measurement. The particle configuration then remains associated with this choice of configuration space wavepacket during the subsequent time evolution. Essentially, once the particle configuration is localized within a world, it cannot cross to another world. This is because of the disjointness of the spatial supports of the configuration space wavepackets (or alternatively, the corresponding disjointness of the branches in Track-1).

**Subsystem Compatible interpretation.** The time evolution in the Subsystem Compatible interpretation has the following properties:

1. In the Subsystem Compatible interpretation, the particle configuration begins in agreement with the standard de Broglie-Bohm interpretation. But rather than have the particle configuration propagate along the branching structure of Track-1, each particle propagates along a projection of this structure onto three-dimensional space. The projection of Track-1 onto the three-dimensional space coordinates of a single particle gives the spacetime diagram for the marginal probability distribution of this particle.

2. Provided the projection remains branching and disjoint at the current location of the particle, the correlations in the joint probability density corresponding to particle measurement are preserved during the particle's propagation. This is because the particle's motion is equivalent to propagation according to Track-1 for this section of time evolution.

3. There are, however, times when the projected structure contains loops which are not present in the configuration space structure of Track-1. This occurs whenever the disjoint wavepackets, corresponding to different measurement outcomes, overlap one another again in three-dimensional space. These wavepackets can overlap in three-dimensional space, but remain separate in configuration space. In the event that this occurs, the particle in question can essentially cross from one world to another.[5]

4. However, there are several mechanisms which ensure the quantum mechanical predictions for observed phenomena are produced despite some of the particles being able to cross worlds.

    (a) *Argument of likelihood:* Once the single-particle wavepackets have moved to regions of disjoint spatial support to be measured, it is very unlikely that they will cross paths again. Typically, they will either become attracted to other physical matter or be become dispersed.

    (b) *Argument of macroscopic experience:* Even if some of the particles have wavepackets that do encounter one another again in three-dimensional space, on aggregate, the effect of these

---

[5]Note that our use of the terminology 'cross worlds' is slightly misleading. There is only one world in the ontology. But the particle can become associated with the three-dimensional space wavepacket which is part of a different world in the standard de Broglie-Bohm configuration space description.

particles crossing worlds is irrelevant. There is a general entropic drive toward wavepackets to remain in disjoint regions of three-dimensional space. Therefore on aggregate, bulk collections of particles do not display any deviations from the quantum predictions. Now since all observations are performed at the classical level (i.e. reading the classical detector states corresponding to measurement) what we experience are the outcomes for bulk collections of particles which match the quantum predictions at all times.

- (c) *Argument of eventual irrelevance:* All loops in the projected branching structure eventually terminate into input modes which match the quantum predictions. This is because the initial states of particles are always separable. The measurement results for separable states are given by the marginal probabilities, and so are necessarily produced by the particles propagating according to the de Broglie-Bohm guidance equations of single-particle subsystems.

5. The trajectories of an ensemble of these particle configurations also form a branching topological structure in configuration spacetime. We will call this structure *Track-2*. Track-1 and Track-2 contain a lot of similarities in their branching structure (point 2). Guidance along either track, due to these similarities, allows both structures to produce a sample from the quantum predictions for macroscopic phenomena (points 4a, 4b, 4c). However, as stated above, Track-2 contains unobservable looped deviations which are not present in Track-1 (point 3). These deviations are essentially appendages to Track-1, such that the topological structure is compatible with retrocausal motion according to single-particle guidance equations.

## 3.4 Advantages of the Subsystem Compatible interpretation

We now summarize the main advantages of the Subsystem Compatible interpretation:

1. The interpretation describes the quantum mechanical predictions for observable phenomena in a local way.
2. The interpretation is compatible with a local realist description of quantum subsystems.
   - (a) The interpretation satisfies locality and counterfactual definiteness of the marginal probabilities.
   - (b) The interpretation is sufficiently constrained to be compatible with the existence of the local, forwardly causal description of the marginal probabilities provided in Section 2. The interpretation has this description of the marginals embedded within a broader ontology for the full quantum predictions.
3. The interpretation is minimally retrocausal, using only retrocausal particles rather than retrocausal waves. It can also be recast as a superdeterminstic forwardly causal interpretation if preferred.
4. The interpretation avoids closed causal loops since the particle has no influence upon the dynamics of the wavefunction. Therefore it exploits this property which has been criticized in the standard de Broglie-Bohm interpretation.

## 3.5 The Subsystem Compatible Interpretation (superdeterministic version)

Because the Subsystem Compatible interpretation is local and deterministic, it is easy to recast it as a forwardly causal interpretation of quantum mechanics. We outline the procedure to do so.

**Preliminary step:**
1. Specify the final particle configuration in accordance with the joint probability distribution $|\Psi(\vec{x}_1,\vec{x}_1,...,\vec{x}_n,t_f)|^2$.
2. Propagate this particle configuration backward in time according to the Subsystem Compatible interpretation (retrocausal version). This produces an initial particle configuration which corresponds to the entire motion undertaken.

**Superdeterministic interpretation:**
1. Choose this previously determined initial particle configuration as an initial boundary condition.
2. Propagate the particle configuration forward in time according to the de Broglie-Bohm guidance equations of single-particle subsystems. The particle configuration will undergo the exact same motion as in the retrocausal version of the theory. However since the guidance equations of single-particle subsystems are local and deterministic, neither non-locality nor retrocausality are required to explain the quantum predictions.

The forwardly causal version of the Subsystem Compatible interpretation is quite interesting in the context of Bell's theorem. It exploits the superdeterminism loophole to evade the theorem. The particles of the detector must arrange themselves in the correct way so as to elicit the correct correlations between the particles of the Bell pair.

## 3.6 Summary of results

Converting the de Broglie-Bohm interpretation into a local retrocausal interpretation cannot be achieved by simply specifying the final particle configuration and propagating backward in time. The many-body guidance equations are non-local even in the reverse direction of time. However the final particle configuration can be propagated according to de Broglie-Bohm guidance equations of single-particle subsystems. These guidance equations are local, provided the local sum-over-paths interpretation is used to describe them.

In the standard de Broglie-Bohm interpretation, observed phenomena are encoded onto the branching topological structure which is formed in configuration spacetime by an ensemble of particles over their motion. This branching structure can be modified slightly, such that it is compatible with the motion of retrocausal particles propagating according to the guidance equations of single-particle subsystems. Several mechanisms ensure the elements of the branching structure related to observable phenomena are preserved, despite these modifications.

# 4 Conclusion

This series of papers has been concerned with the ontology of quantum subsystems. Paper one has demonstrated that the standard approaches to quantum mechanics are problematic with respect to their description of spatially entangled subsystems. Paper two has established that a local sum-over-paths interpretation of quantum subsystems can resolve this issue. This paper has demonstrated that a local interpretation of quantum subsystems is compatible within a broader ontological framework for the full quantum predictions.

Bell's theorem leaves open three main loopholes by which a local description of quantum mechanics can be obtained. These loopholes are retrocausality, superdeterminism and a rejection of counterfactual definiteness respectively. The Subsystem Compatible interpretation primarily exploits the retrocausality loophole. The interpretation describes retrocausal de Broglie-Bohm particles being driven in the opposite temporal direction to the wavefunction by the guidance equations of single-particle subsystems. As well as describing the full set of observable quantum mechanical phenomena, the interpretation is constrained to satisfy locality and counterfactual definiteness of the marginal probabilities. Table 1 compares the de Broglie-Bohm, Copenhagen and Subsystem Compatible interpretations with respect to the questions of locality and counterfactual definiteness.

| A COMPARISON OF INTERPRETATIONS | LOCALITY? | COUNTERFACTUAL DEFINITENESS? |
| --- | --- | --- |
| MANY-BODY DE BROGLIE-BOHM | No. | Yes, for all phenomenon. |
| COPENHAGEN INTERPRETATION | No. | Yes, for observable phenomenon only. |
| SUBSYSTEM COMPATIBLE INTERPRETATION | Yes. | Yes, for observable phenomenon. Yes, for subsystems. |

Table 1: Comparison of interpretations with respect to locality and counterfactual definiteness.

The Subsystem Compatible interpretation also has a quite unique ontology. The general aim has been to constrain the interpretation of the quantum mechanical joint probabilities by requiring concordance with a local, forwardly causal interpretation of the marginal probabilities. The interpretation achieves this objective, since it is built directly from such an interpretation of the marginals. It contains a local realist description of single-particle subsystems as a subset of its wider explanatory framework. This fact is made particularly evident in the superdeterministic formulation of the interpretation. The superdeterministic formulation eliminates the element of retrocausality, leaving only the local realist description of subsystems. Since the true nature of entanglement correlations remains unknown, it may be wise to constrain our best physical theories by the principle of local realism to the maximum extent that nature will allow.

# A  Generalizing the local sum-over-paths

In this section, we describe how the local sum-over-paths approach can be generalized to describe both spatial and spin entanglement. We want to account for state vectors from the Hilbert space $\mathcal{H}_{\text{Space A}} \otimes \mathcal{H}_{\text{Spin A}} \otimes \mathcal{H}_{\text{Space B}} \otimes \mathcal{H}_{\text{Spin B}}$ i.e. state vectors which are tensor products of both spatial and spin degrees of freedom. The probability amplitudes for particle A to occur in mode $j$ and spin state $a$, and particle B to end in mode $k$ and spin state $b$ can be given by the following matrix expansion of the unitary $\mathbb{U}^{(n)}$:

$$\langle j,a,k,b|\mathbb{U}^{(n)}|0\rangle = \langle j,a,k,b|\mathbb{A}^{(n)} \otimes \mathbb{B}^{(n)} \mathbb{AB}^{(n)}...\mathbb{A}^{(2)} \otimes \mathbb{B}^{(2)} \mathbb{AB}^{(2)} \mathbb{A}^{(1)} \otimes \mathbb{B}^{(1)} \mathbb{AB}^{(1)}|0\rangle, \quad (A.1)$$

where the single particle unitaries $\mathbb{A}^{(t)}$ and $\mathbb{B}^{(t)}$ and controlled phase gates $\mathbb{AB}^{(t)}$ act on kets from the appropriate tensor product of Hilbert spaces. The local sum-over-paths approach works analogously to the spatial mode case, but we now need to decompose in terms of paths through the combined spatial mode$\times$ spin space. Furthermore, we also need to perform a classical sum over the spin states of particle A. Explicitly, the marginal probabilities for particle A to be in mode $j$ are:

$$\begin{aligned} P(A=j) &= \sum_a \sum_b \sum_k |\langle j,a,k,b|\mathbb{U}^{(n)}|0\rangle|^2 \\ &= \sum_a \sum_b \sum_k |\sum_{P(j,a,n)} \langle k,b|\mathbb{U}_P^{(n)}|0\rangle|^2 \\ &= \sum_a \sum_b \sum_k \sum_{P(j,a,n)} \sum_{Q(j,a,n)} \langle 0|\mathbb{U}_P^{(n)\dagger}|k,b\rangle \langle k,b|\mathbb{U}_Q^{(n)}|0\rangle, \end{aligned} \quad (A.2)$$

where $\sum_{P(j,a,n)}$ is a sum over space $\times$ spin paths $P$ of particle A, leading to spatial mode $j$ and spin state $a$ at layer $n$ of the circuit. To generalize to the case of describing the marginal probability density, divide the position space up into a discrete lattice of infinitesimal spatial modes with kets $|x_1\rangle$. Then the analysis carries through as in the case of spatial modes. We obtain in this case:

$$\begin{aligned} P(A=\vec{x}_1) &= \sum_a \sum_b \sum_k |\langle \vec{x}_1,a,\vec{x}_2,b|\mathbb{U}^{(n)}|0\rangle|^2 \\ &= \sum_a \sum_b \sum_k |\sum_{P(\vec{x}_1,a,n)} \langle \vec{x}_2,b|\mathbb{U}_P^{(n)}|0\rangle|^2 \\ &= \sum_a \sum_b \sum_k \sum_{P(\vec{x}_1,a,n)} \sum_{Q(\vec{x}_1,a,n)} \langle 0|\mathbb{U}_P^{(n)\dagger}|\vec{x}_2,b\rangle \langle \vec{x}_2,b|\mathbb{U}_Q^{(n)}|0\rangle, \end{aligned} \quad (A.3)$$

where the event that particle A is located at $\vec{x}_1$ means that this particle is located at the lattice site of the spatial mode $|\vec{x}_1\rangle$.